\documentclass[twocolumn,letter]{jpsj2} 
\title{Effects of Randomness on the Field-Induced Phase Transition in the S=1 Bond-Alternating 
Spin Chain NTENP}

\author{Shinichi \textsc{Matsubara}$^{1}$, Katsuaki \textsc{Kodama}$^{1}$\thanks{Present address:
Neutron Science Research Center, JAERI.}, Masashi \textsc{Takigawa}$^{1}$\thanks{E-mail address: 
masashi@issp.u-tokyo.ac.jp}, and Masayuki \textsc{Hagiwara}$^{2}$}

\inst{$^{1}$Institute for Solid State Physics, University of Tokyo, Kashiwa, Chiba 277-8581, Japan \\
$^{2}$KYOKUGEN, Osaka University, Toyonaka, Osaka 560-8531, Japan}

\abst{We report novel effects of randomness in the S=1 bond-alternating antiferromagnetic 
chain compound with a dimer-singlet ground state 
[Ni(N,N'-bis(3-aminopropyl)propane-1,3-diamine($\mu$-NO${}_2$)]ClO${}_4$   
abbreviated as NTENP. The $^{15}$N NMR spectra develop a continuum with 
sharply peaked edges at low temperatures, indicating an inhomogeneous staggered 
magnetization induced by magnetic field.  We attribute this to random anisotropic interactions 
due to disorder of NO$_2$ groups in the chains.  The field-induced antiferromagnetic 
transition exhibits remarkably anisotropic behavior.  We propose that a field-induced 
incoherent magnetization is transformed into a coherent antiferromagnetic moment 
with spatially fluctuating amplitude.}

\kword{NTENP, random effects, field-induced magnetic transition, nuclear magnetic resonance}

\begin{document}
\maketitle

Among various quasi-one-dimensional quantum antiferromagnets, 
S=1 bond-alternating chains expressed by the Hamiltonian 
\begin{equation}
\mathcal{H}_0 =  J \sum_i \left( \mib{S}_{2i-1}\cdot \mib{S}_{2i}+ 
\alpha \mib{S}_{2i} \cdot \mib{S}_{2i+1} \right)  
\label{H0}
\end{equation}
has an interesting feature that two distinct singlet phases appear as the bond alternating 
ratio $\alpha$ varies from zero to one~\cite{affleck871}. 
The dimer-singlet phase is realized for $\alpha<\alpha _c$=0.588~\cite{kohno981}, 
while the Haldane phase with the valence-bond-solid states appears for $\alpha>\alpha _c$. 
The triplet excitations have a finite energy gap except at $\alpha=\alpha_{c}$~\cite{hagiwara981}.  
 
The triplet excitations are split by a magnetic field ($\mib{H}$) .  The energy of the 
lowest branch is reduced with the field and vanishes at a critical field $H_{c}$.  For a 
finite interchain coupling, the Bose-Einstein condensation (BEC) of the triplets occurs 
generally for $H > H_{c}$~\cite{affleck911,nikuni001}.  Since the bosons are created 
by the transverse spin operator at the antiferromagnetic wave vector, the BEC is 
equivalent to an antiferromagnetic (AF) order perpendicular to the magnetic field. 
While this is a generic scenario, the field-induced AF transitions in real materials 
show a variety of behavior depending on various perturbations such as 
anisotropy, frustrating interactions, and disorder, which are not included in eq.~(\ref{H0}).   

The single ion anisotropy energy in a uniaxial S=1 system is given by 
$D \sum_{i} S_{iz}^{2}$.  For $H \parallel z$ the AF order perpendicular to 
the field has the $XY$-symmetry, while for $H \perp z$ the direction of the AF  
moment is uniquely fixed, resulting in the Ising symmetry. Other types of anisotropy 
are allowed for crystals with low symmetry, for example, the Zeeman interaction 
involving anisotropic $g$-tensors $\mib{H} \cdot \sum_{i} \mib{g}_{i} \cdot \mib{S}_{i}$
and the Dzyaloshinskii-Moriya (DM) interaction~\cite{dzyaloshinskii, moriya} 
$\sum_{i} \mib{d}_{i} \cdot \left(  \mib{S}_{i} \times \mib{S}_{i+1} \right)$.
If such interactions contain a staggered 
part with sign alternation, $\mib{g}_{i} = \mib{g}^{u} + (-1)^{i}\mib{g}^{s}$ or  
$\mib{d}_{i} = \mib{d}^{u} + (-1)^{i}\mib{d}^{s}$, application of a uniform magnetic field 
leads to an effective staggered field $(-1)^{i}\mib{H}_{s}$ with $\mib{H}_{s}\propto 
\mib {g}^{s} \cdot \mib{H}$ or $\mib{H}_{s} \propto \mib{d}_{s} \times \mib{H}
/J$~\cite{affleck041}.  The staggered field produces a staggered 
magnetization even for $H < H_{c}$.  It also mixes the singlet with the triplets,  
preventing the gap from closing at $H_{c}$~\cite{chiba911, mitra941}.  
These phenomena have been observed, for example,  in the Haldane chain compound 
Ni(C$_{2}$H$_{8}$N$_{2}$)$_{2}$NO$_{2}$(ClO$_{4}$) 
(NENP)~\cite{chiba911,mitra941}.  (Similar phenomena have 
been observed also in the 2D frustrated dimer spin system 
SrCu$_{2}$(BO$_{3}$)$_{2}$~\cite{miyahara041,kodama051}. )   
Such diversity of the quantum critical phenomena in real materials 
is a fascinating aspect of the physics of spin systems in magnetic fields.   

\begin{figure}[b]
\begin{center}
\includegraphics[width=5cm]{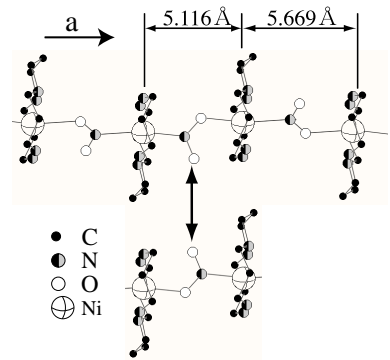}
\end{center}
\caption{The structure of a magnetic chain in NTENP. Two types of 
Ni-NO$_2$-Ni bonds related by inversion occur randomly.  
Note that individual bonds do not have the inversion symmetry.}
\label{structure}
\end{figure}
[Ni(N,N'-bis(3-aminopropyl)propane-1,3-diamine($\mu$-NO$_2$)]ClO$_2$
abbreviated as NTENP with the chemical formula 
Ni(C$_{9}$H$_{24}$N$_{4}$)(NO$_{2}$)(ClO$_{4}$)
is an example of S=1 bond-alternating chain systems.  The chains, separated by 
ClO$_{4}^{-}$ anions, run along the $a$-axis of the triclinic $P\bar{1}$ structure, 
where the Ni$^{2+}$ ions are connected via nitrito groups (NO$_{2}^{-}$) 
with alternation of short and long bonds (Fig.~\ref{structure})~\cite{escuer971}.  
The results of susceptibility, magnetization~\cite{narumi011}, and neutron inelastic 
scattering~\cite{zheludev041}for NTENP, as well as the ESR data for Zn-doped 
NTENP~\cite{narumi011} established that NTENP belongs to the dimer-singlet 
phase with the exchange parameters $\alpha$=0.45, $J$=54~K and the nearly
uniaxial single-ion anisotropy $D$=13.6~K.  The latter is consistent with the critical fields 
for the onset of magnetization, $H_{c}$=9.3~T for $\mib{H} \parallel \mib{a}$ and  
$H_{c}$=12.4~T for $\mib{H} \perp \mib{a}$~\cite{narumi011}, and the triplly split 
excitation gap energies, 1.06, 1.15, and 1.96~meV, observed by the neutron scattering 
experiments~\cite{hagiwara051}.  
The specific heat shows a pronounced sharp peak at low temperatures 
for $H > H_{c}$~\cite{tateiwa031}, indicating a well developed long range AF 
order.  The AF order, which was confirmed by neutron diffraction~\cite{hagiwara051}, 
can be interpreted as the BEC of triplets. NTENP also has a unique structural feature~\cite{escuer971}. 
A nitrogen atom on each bond can be located next to either the left or the right 
Ni atom randomly with equal probability (Fig.~\ref{structure}). 
However, no effect of randomness has been reported so far.  

In this paper, we report observation of a field-induced inhomogeneous staggered 
magnetization in a single crystal of NTENP by nuclear magnetic resonance (NMR) 
experiments on $^{15}$N (nuclear spin $I$=1/2) and $^{35}$Cl ($I$=3/2) nuclei.  
We attribute this to the structural randomness, which also causes highly anisotropic 
behavior of the field-induced magnetic transition.  Only the N atoms in the NO$_{2}^{-}$
groups were replaced by $^{15}$N isotope.  The NMR spectra were obtained 
from the Fourier transform of the spin-echo signal.  The nuclear relaxation rate 
1/$T_{1}$ was measured by the inversion recovery method. 

\begin{figure}[t]
\begin{center}
\includegraphics[width=8.5cm]{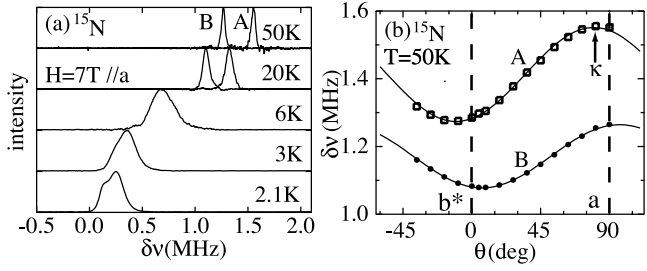}
\end{center}
\caption{(a) The $^{15}$N NMR spectra at H=7~T along the $a$-axis at several temperatures.  
(b) Angular dependence of $\delta\nu$ at $T$=50~K for the 
field in the $ab^{*}$-plane.  A and B denotes the different sites. The lines show 
sinusoidal curves with the period of 180$^{\circ}$.} 
\label{15NhighT}
\end{figure}
The $^{15}$N-NMR spectra are shown in Fig.~\ref{15NhighT}(a) 
at various temperatures for the field of 7~T along the $a$-axis.  Here the 
signal intensity is plotted against the frequency shift 
$\delta\nu = \nu - \gamma_{N}H$, where $\nu$ is the signal frequency and 
$\gamma_{N}$ is the nuclear gyromagnetic ratio (4.3142~MHz/T for $^{15}$N 
and 4.1717~MHz/T for $^{35}$Cl).  The signal frequency of a nucleus is 
proportional to the sum of the external field and the time-averaged magnetic 
hyperfine field due to Ni spins, 
$\nu = \gamma_{N}(|\mib{H} +  \langle \mib{H}^{hf} \rangle|)$ with 
$\langle \mib{H}^{hf} \rangle = \sum_{i}\mib{A}_{i} \cdot   \mib{g}
\cdot \langle \mib{S}_{i} \rangle$, where $\mib{A}_{i}$ 
is the hyperfine coupling tensor between the nucleus and the nearby
spin $\mib{S}_{i}$ and $\langle \rangle$ stands for the time-average.  Since 
$H \gg |\langle \mib{H}^{hf}\rangle|$, the frequency shift is given by the projection of 
$\langle \mib{H}^{hf} \rangle$ along the 
external field direction, $\delta\nu = \mib{H} \cdot \langle \mib{H}^{hf} \rangle /H$. 
The NMR spectra then represent the distribution of $\delta\nu$.      

Two NMR lines are observed at 50~K, corresponding to the distinct  
N sites on the short and the long bonds with different hyperfine 
coupling tensors to the uniform magnetization $\mib{A}_{u}=\sum_{i}\mib{A}_{i}$.  
The A-site with the larger shift is likely to be on the short bond.  
The shifts at both sites are reduced to nearly zero at low temperatures, where two lines 
overlap (Fig.~\ref{15NhighT}(a)), in a way similar to the susceptibility.  
However, the line width \textit{increases} with decreasing temperature, 
indicating inhomogeneous distribution of magnetization 
at low temperatures.  We discuss the line width in more detail below. 

In Fig.~\ref{15NhighT}(b), $\delta\nu$ at $T$=50~K is plotted as a function of the 
angle $\theta$ between the $b^{*}$-axis and the field in the $ab^{*}$-plane. 
For the A-site, $\delta\nu$ has a maximum at $\theta$=80$^{\circ}$ 
(denoted as $\kappa$).  The global maximum in the plane containing 
the $\kappa$-axis and perpendicular to the $ab^{*}$-plane occurs 
at 4$^{\circ}$ from the $\kappa$-axis (defined as $\xi$, not shown), 
which should then be one of the principal axes of $\mib{A}_{u}$. 
The standard $K$-$\chi$ analysis 
yields the principal value as $A_{u}^{\xi \xi}$=4.2~T/$\mu_{B}$.  
Other principal values are obtained as 2.9 and 3.1~T/$\mu_{B}$.  
Thus $\mib{A}_{u}$ is approximately uniaxial along the Ni-N bond 
direction with a large isotropic part.  Since these values are an order of 
magnitude larger than the dipolar field, the major source of the hyperfine 
field should be the spin density on the 2s and the 2p states of N sites 
transferred from the d states of the neighboring Ni sites.  
We expect the same mechanism for the B-sites based on the similar magnitude and 
$\theta$-dependence of  $\delta\nu$ in the $ab^{*}$-plane, 
although the complete angular dependence has not been measured.   

\begin{figure}[t]
\begin{center}
\includegraphics[width=8cm]{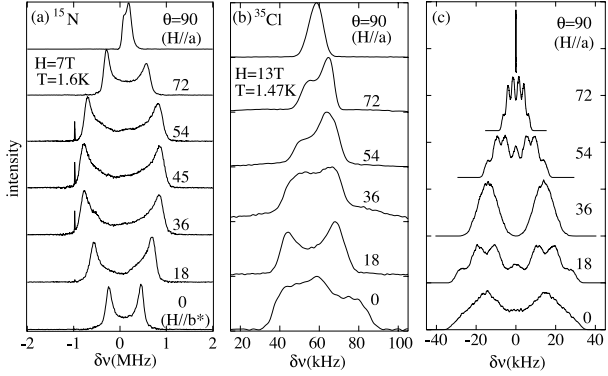}
\end{center}
\caption{Angular dependence of the NMR spectra for the field in the $ab^{*}$-plane. 
(a) $^{15}$N spectra at $T$=1.6~K and $H$=7~T. 
The sharp peak near -1~MHz is due to $^{35}$Cl nuclei.  (b) $^{35}$Cl spectra at $T$=1.4~K 
and $H$=13~T. (c) The simulated $^{35}$Cl spectra for the model described in the text.} 
\label{lowTspec}
\end{figure}

To understand the mechanism for the line broadening at low temperatures, 
we examined angular dependence of the line shape at 1.6~K for the field in the 
$ab^{*}$-plane as shown in Fig.~\ref{lowTspec}(a).  When the field is 
away from the $a$- or $b^{*}$-axes, the spectra develop a double peak  
structure with a continuum and sharp edges on both sides.  The largest peak splitting  
at $\theta$=45$^{\circ}$ ($\sim$2~MHz) is about two orders of magnitude 
larger than the width of individual lines at 50~K, while the bulk magnetization at 1.6~K is 
an order of magnitude \textit{smaller} than the value at 50~K. Such broadening must be   
due to continuous distribution of the hyperfine field with a large weight near the extremal  
values.  The nearly symmetric line shape with respect to $\delta\nu$=0 indicates that
its origin is a local staggered magnetization with spatial inhomogeneity.  
We observed that the line width changes linearly with the field, therefore, the 
staggered magnetization must be induced by the external field. 

The angular dependence of the line shape strongly suggests that the local staggered 
magnetization is perpendicular to the field.  If it were 
parallel to the field, the peak splitting should be largely independent of the field direction
since the hyperfine coupling tensor is dominantly isotropic.  On the other hand, 
splitting due to perpendicular magnetization is caused only by the anisotropic 
part of the hyperfine coupling.  The more the field deviates from the principal 
axes of the hyperfine coupling tensor, the larger should become the splitting.  
This is precisely what we observed.  Upon closer inspection, however, we note that 
the spectrum becomes narrowest for $\mib{H} \parallel \mib{a}$, even though the $a$-axis 
deviates slightly from the principal axes.  In contrast, clear splitting is observed for 
$\mib{H} \parallel \mib{b}^{*}$ while the $b^{*}$-axis deviates from the closest principal 
axis by a similar amount as the $a$-axis. This suggests that the magnitude of the 
staggered magnetization varies with the field direction and becomes smallest for 
$\mib{H} \parallel \mib{a}$.    

To confirm this point, we examined the $^{35}$Cl NMR spectra.  
Since a Cl nucleus is coupled to many Ni spins via dipolar interaction, 
the hyperfine field should be less sensitive to the direction of the 
random magnetization. The angular variation of the $^{35}$Cl NMR 
spectra of the central transition ($I_{z} = -1/2 \leftrightarrow 1/2$) 
at $T$=1.47~K are shown in Fig.~\ref{lowTspec}(b).   Although 
the field (13~T) exceed $H_{c}$, no long range AF order occurs 
at this temperature~\cite{tateiwa031}. The line width shows a 
minimum for $\mib{H} \parallel \mib{a}$ and a maximum for 
$\mib{H} \parallel \mib{b}^{*}$ in contrast to the case of $^{15}$N spectra. 
This supports that the staggered magnetization is minimized for $\mib{H} \parallel \mib{a}$.  

The field-induced inhomogeneous staggered magnetization observed by NMR 
can be qualitatively understood by the structural disorder mentioned before.    
Since the individual Ni-Ni bond breaks the inversion symmetry, 
the following DM interaction should be allowed
\begin{equation}
\mathcal{H}_{DM}= \sum_{j}\left[ \pm \mib{d}_{1} \cdot  
\left( \mib{S}_{2i-1} \times \mib{S}_{2i} \right) 
\pm \mib{d}_{2}\cdot  \left( \mib{S}_{2i} \times \mib{S}_{2i+1}\right) \right] , 
\label{Hrandom}
\end{equation}
where the first (second) term applies to the short (long) bond 
and the random sign arises from the two configurations of NO$_{2}$ 
groups related by inversion. The structural disorder should cause 
random modulation of the $g$-tensor as well. 
Although this will also lead to a random staggered 
magnetization, how the $g$-tensor depends on bond configurations is not 
obvious and we leave the detailed analysis for future studies.    
  
We first consider the simplest case of isolated dimers ($\alpha$=0) 
with $\mib{d}_{1} \parallel \mib{a}$. The external field induces staggered 
magnetization with the random sign~\cite{affleck041,miyahara041,kodama051} 
$\langle \mib{S}_{2i-1}\rangle - \langle \mib{S}_{2i}\rangle 
= \pm (8g/3J^{2}) \mib{d}_{1} \times \mib{H} =  \pm M_{s}\cos \theta$.  
We have simulated the $^{35}$Cl spectra due to the dipolar field from such 
magnetization with $M_{s}$=0.4$\mu_{B}$ as shown in Fig.~\ref{lowTspec}(c).
Even this simplest model reproduces qualitative feature of the data, 
providing estimation for the size of the staggered magnetization.
However, this does not account for the spectra of $^{15}$N nuclei, 
which should couple dominantly to two neighboring  Ni spins. 
For the B-sites on the weak bonds, the sign change of the staggered 
magnetization leads to a line splitting symmetric about $\delta\nu$= 0. 
Contrary, the hyperfine field at the A-sites on the strong bonds should not 
be altered by the inversion that flips the staggered magnetization of the same
bond. We then expect a superposition of doubly peaked and singly 
peaked spectra, inconsistent with the data. 

However, the interdimer interaction is not weak in NTENP ($\alpha \sim$ 0.5) and 
one AF correlation length should contain several dimers. The random
perturbation (eq.~\ref{Hrandom}) may enhance the AF correlation set by the 
main Hamiltonian (eq.~\ref{H0}) in some regions, but may frustrate 
with it in other regions.  This competition (frustration) will bring spatial fluctuations 
both in the correlation length and in the size of the staggered magnetization.
Such distribution is likely to produce continuous spectra for both 
N sites, although more precise understanding of the $^{15}$N spectra based 
on a realistic model is highly desired.  We emphasize that long range coherence 
should never be established for the field-induced magnetization below $H_{c}$, 
therefore, it cannot be detected by neutron diffraction.   

\begin{figure}[t]
\begin{center}
\includegraphics[width=7cm]{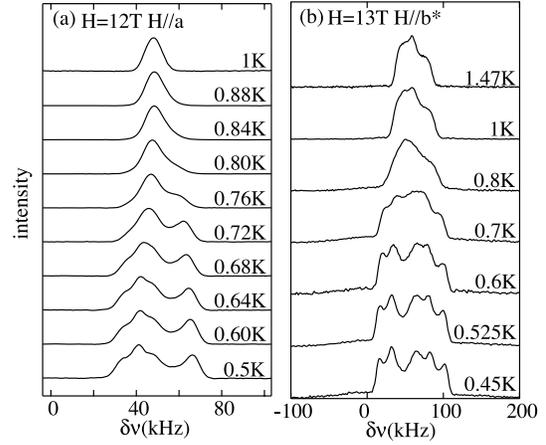}
\end{center}
\caption{Temperature variation of the $^{35}$Cl NMR spectra across the  
of phase transition for (a) $H$=12~T along the $a$-axis and (b) $H$=13~T along 
the $b^{*}$-axis.} 
\label{FIPT}
\end{figure}
We found that such incoherent staggered magnetization has 
profound effects on the field-induced AF transition.  Evolution of 
the $^{35}$Cl NMR spectra upon entering into the AF phase 
by lowering temperature is shown in Fig.~\ref{FIPT} for (a) 
$H$=12~T along the $a$-axis and (b) $H$=13~T along the 
$b^{*}$-axis.  The transition temperature $T_{c}$ can be inferred 
from the specific heat data~\cite{tateiwa031} to be 0.86~K (case a) 
and 0.88~K (case b).  For the case (a), the spectrum gets broadened 
significantly in a relatively narrow temperature range below T$_{c}$ 
(0.84~K $\ge T \ge$0.72~K) and broad double peak 
structure appears at lower temperatures.  If there were no disorder, 
an AF phase transition generally causes NMR lines to split 
and the splitting is proportional to the AF order parameter.  
In the presence of substantial disorder, an AF order results in line 
broadening rather than splitting.  Thus the data for the case (a) are 
consistent with what one would expect for a second order AF 
transition with strong disorder.  For the case (b), however, the 
spectrum has a broad shape already above $T_{c}$
 (Fig.~\ref{lowTspec}b). Only a minor change occurs in a wide 
temperature range below $T_{c}$.  One would not 
have noticed a phase transition from the NMR spectra alone.   

The peculiar behavior for $\mib{H} \parallel \mib{b}^{*}$ can be 
qualitatively understood if one realizes that an incoherent staggered 
magnetization is already developed above $T_{c}$. Below $T_{c}$, 
the long range coherence must be established as indicated by the 
sharp Bragg peaks of neutron diffraction~\cite{hagiwara051}. 
However, there should still be large spatial fluctuations in the magnitude due to 
frustration between the regular AF interactions and the random 
anisotropic interactions.  Since the NMR spectrum is sensitive only to  
local magnetization profile, the states above and below $T_{c}$ may not
be clearly distinguishable by NMR .  More normal behavior is expected 
for $\mib{H} \parallel \mib{a}$ because the staggered magnetization above 
$T_{c}$ is minimum. Usually the alternating DM interaction without
disorder inhibits a phase transition as has been observed in NENP.  
It is interesting that in NTENP the disorder enables the phase transition.  
 
\begin{figure}[t]
\begin{center}
\includegraphics[width=7cm]{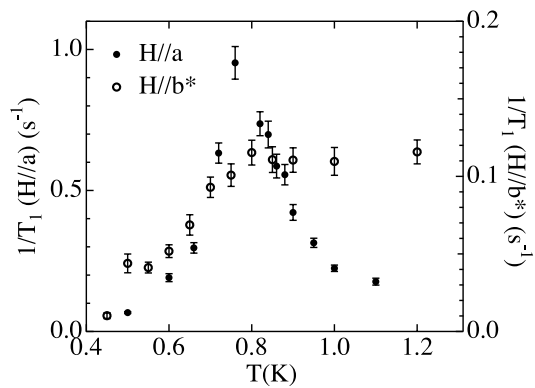}
\end{center}
\caption{The temperature dependence of 1/$T_{1}$ for $^{35}$Cl nuclei for   
$H$=12~T parallel to the $a$-axis (solid circles, left scale) and $H$=13~T parallel to the $b^{*}$-axis
(open circles, right scale).} 
\label{T1}
\end{figure}
The field-induced AF transition exhibits remarkable anisotropy also
in the dynamical properties.  Temperature dependence of the 
nuclear spin-lattice relaxation rate $1/T_1$ for $^{35}$Cl nuclei is 
shown in Fig.~\ref{T1}. We found that the nuclear magnetization follows
an exponential time dependence after it recovers to about 50 \% of
the equilibrium value, even though the initial recovery was non-exponential. 
For $\mib{H} \parallel \mib{a}$, 1/$T_{1}$ 
shows a huge enhancement near $T_{c}$ due to slowing down 
of spin fluctuations.  It seems unlikely that the enhancement of 1/$T_{1}$ 
over such a wide temperature range is related to 3D ordering.  
We speculate that it may be due to enhanced 1D fluctuations 
in the Luttinger liquid regime, since the field direction maintains the 
approximate XY-symmetry. 

In contrast, no enhancement was observed for
$\mib{H} \parallel \mib{b}^{*}$.  1/$T_{1}$ stays nearly constant down to $T_{c}$, 
below which it is depressed rapidly.  This behavior could arise from two reasons.
First, the field-induced AF order has the Ising symmetry for 
$\mib{H} \parallel \mib{b}^{*}$, leading to a narrower critical temperature 
range. However, this alone would not eliminate the enhancement of 1/$T_{1}$ 
completely.  Second, the spin fluctuations above $T_{c}$ should be 
suppressed significantly by the field-induced random magnetization.  
We suppose that the incoherent magnetization above $T_{c}$ is transformed 
into a coherent AF moment below $T_{c}$ by the reversal of magnetization 
of domains whose size is set by the locally varying correlation length. 
The constant behavior of $1/T_{1}$ above $T_{c}$
indicates that this process does not involve slowing down of fluctuations.  
Nevertheless, the phase transition can be identified clearly by the sudden 
drop of 1/$T_{1}$ below 0.8~K.  We notice that the temperature for the 
anomaly of 1/$T_{1}$ (0.76~K for $\mib{H} \parallel \mib{a}$ and 0.8~K for 
$\mib{H} \parallel \mib{b}^{*}$) is slightly lower than $T_{c}$ determined by 
specific heat.  We do not understand the reason at moment.  Detailed NMR 
measurements of the phase diagram and comparison with other experimental results 
remain to be done in future.   
  
In conclusion, anomalous line broadening of the $^{15}$N and $^{35}$Cl NMR 
spectra in NTENP provide evidence for field-induced incoherent staggered magnetization
due to structural disorder.  We propose that the anisotropic response of this  
random magnetization significantly affects both the static and the dynamic behavior 
of the field-induced antiferromagnetic transition.  The phenomena found in NTENP 
resemble the "random field effects" studied extensively a few decades 
ago~\cite{aharony861} in that both are caused by frustration between the 
regular AF interactions and the random magnetic field.  
In NTENP, however, the random interaction appears to have no effects on the singlet 
ground state at zero-field. The magnetic field induces \textit{both} the AF order 
\textit{and} the effective random field.  We believe that this is a new type of quantum phase
transition involving randomness and calls for serious theoretical investigations.        
We would like to thank J. Yamaura for help in crystal orientation by X-ray measurements.

\end{document}